\documentclass[12pt]{iopart}

\usepackage{iopams}  

\usepackage{graphicx}
\usepackage{dcolumn}
\usepackage{bm}
\newtheorem{Thm}{Theorem}
\newtheorem{Lem}[Thm]{Lemma}
\newtheorem{Cor}[Thm]{Corollary}

\newtheorem{Def}[Thm]{Definition}

\begin{document}

\title[
Theorems on ground-state phase transitions 
in Kohn-Sham models 
]{
Theorems on ground-state phase transitions 
in Kohn-Sham models given by the Coulomb density functional 
}

\author{Koichi Kusakabe, Isao Maruyama}

\address{
Graduate School of Engineering Science, Osaka University, 
1-3 Machikaneyama-cho, Toyonaka, Osaka 560-8531, Japan}
\ead{kabe@mp.es.osaka-u.ac.jp}

\begin{abstract}
Some theorems on derivatives of the Coulomb density functional 
with respect to the coupling constant $\lambda$ are given. 
Consider an electron density $n_{GS}({\bf r})$ given by a ground state. 
A model Fermion system with the reduced coupling constant, $\lambda<1$, 
is defined to reproduce $n_{GS}({\bf r})$ and the ground state energy. 
Fixing the charge density, 
possible phase transitions as level crossings detected 
in a value of the reduced density functional 
happen only at discrete points along the $\lambda$ axis. 
If the density is $v$-representable also for $\lambda<1$, 
accumulation of phase transition points is forbidden 
when $\lambda\rightarrow 1$. 
Relevance of the theorems for the multi-reference density functional 
theory is discussed. 
\end{abstract}

\pacs{
02.30.Sa, 
31.15.ec, 
71.15.Mb 
}
\submitto{\JPA}
\maketitle

\section{Introduction}

The density functional theory (DFT)\cite{Hohenberg-Kohn,Kohn-Sham} 
is one of successful frameworks in the theory of electron systems. 
For a present standard scheme of this theory,
the universal energy density functional $F[n]$ 
defined by the constrained minimization 
method\cite{Levy79,Levy82,Parr_Yang,Driezler_Gross} 
is used as a key quantity. 
Lieb has analyzed $F[n]$, which is often cited as 
the Levy-Lieb functional.\cite{Lieb83} 
This functional connects a Fermion density $n({\bf r})$ 
to a variational energy of an electron wave function $\Psi$ 
showing the same density. 
Correspondence from $n({\bf r})$ to this minimizing state 
vector $|\Psi\rangle$ is shown to exist,\cite{Lieb83} 
which is a basic principle in the density functional theory.

A continuous modification of the Coulomb operator 
appearing in the definition 
of $F[n]$ is one of relevant techniques to analyze 
this functional.\cite{Gunnarsson_Lundqvist,Langreth_Perdew,Parr_Yang,Driezler_Gross}  
The idea is similar to the ordinal perturbation theory of electron 
systems.\cite{Electron-gas,Fetter} 
To have an exact expression 
of the so-called exchange-correlation functional, 
people fixed the charge density and considered 
continuous reduction of the coupling constant 
by multiplying a factor $\lambda$ 
in the energy density functional.\cite{Gunnarsson_Lundqvist,Langreth_Perdew}

We re-analyze this modified functional, 
which is called $F_\lambda[n]$ in this article, to consider 
phase transition points appearing 
in a generalized Kohn-Sham scheme.\cite{Kusakabe-2001} 
We have two main purposes for this analysis. 
The first purpose is to study a parameter differentiability 
of this $\lambda$-modified $F[n]$. 
Fixing a density at an $N$-representable density in ${\cal I}_N$,\cite{Lieb83} 
the parameter derivative of $F_\lambda[n]$ is shown to be well-defined. 
Namely, existence of state vectors giving 
Dini's derivatives is shown. 
As the second purpose, a condition on phase transitions 
around the $v$-representable density is 
derived from the theorems on $F_\lambda[n]$. 
Existence of level-crossing points, {\it i.e.} phase transition points, 
is known. We address existence of an $\varepsilon$-vicinity 
around the true solution of the many-electron system. 
In this finite region, no level crossing point is found 
provided that the density of the Coulomb system is also 
$v$-representable with a model given by $F_\lambda[n]$. 
The definition of the model is given in section \ref{Kohn-Sham-minimization}. 
Relevance of this $\varepsilon$-vicinity 
for a general Kohn-Sham scheme is discussed in 
section \ref{Summary_conclusions}.

\section{The setup of the problem}
\label{Setup}

We consider a static state of a material. 
The number of electrons $N$ is fixed to be a finite integer. 
We apply the Born-Oppenheimer approximation (BOA), in which 
motion of nuclei is separated from the motion of electrons. 
To find a stable state, 
we consider a classical state of nuclei and 
fix the coordinates of nuclei, whose number is $M$. 
Motion of the electron system is determined for this classical 
configuration of nuclei, which give 
a static electric potential for electrons. 

Let $|0\rangle$ be the electron vacuum. 
The electronic state in an external scalar potential 
is symbolically written by a state vector $|\Psi\rangle$. 
This electron state is described by Fermion field operators, 
$\psi^\dagger_{\sigma}({\bf r})$ and 
$\psi_{\sigma}({\bf r})$, satisfying the canonical anti-commutation relations. 
\[\left[\psi^\dagger_{\sigma}({\bf r}), \psi_{\sigma'}({\bf r}') \right]_+ 
= \delta({\bf r}-{\bf r}') \delta_{\sigma,\sigma'}.\]
Considering a position basis 
$|\left\{{\bf r}_i,\sigma_i\right\}\rangle 
= \prod_{i=1}^N \psi^\dagger_{\sigma_i}({\bf r}_i)|0\rangle$, 
we have a wave function as an inner product between this position 
vector and $|\Psi\rangle$. 
\begin{equation}
\Psi(\left\{{\bf r}_i,\sigma_i\right\}) 
= \langle \left\{{\bf r}_i,\sigma_i \right\}|\Psi\rangle .
\end{equation}

The number operator $\hat{n}({\bf r})$ probing existence of an electron 
at a point ${\bf r}$ is defined as 
\begin{equation}
\hat{n}({\bf r}) \equiv \sum_\sigma 
\psi^\dagger_{\sigma}({\bf r}) \psi_{\sigma}({\bf r}).
\end{equation}
The kinetic energy operator $\hat{T}$ for electrons is assumed to be, 
\begin{equation}
\hat{T}=-\frac{\hbar^2}{2m} \int \! d^3r\, \sum_{\sigma} 
\lim_{{\bf r}' \rightarrow {\bf r}} 
\psi^\dagger_{\sigma}({\bf r}') \Delta_{\bf r} \psi_{\sigma}({\bf r}) ,
\end{equation}
with the electron mass $m$. 

Owing to BOA, we have an external scalar potential 
$v_{\rm ext}({\bf r})$ given by the charge of nuclei. 
Using position vectors ${\bf R}_I$ ($I=1,\cdots,M$) of fixed nuclei, 
the potential $v_{\rm ext}({\bf r})$ is given as, 
\begin{equation}
v_{\rm ext}({\bf r}) = - \sum_{I=1}^M \frac{Z_I e^2}{|{\bf R}_I-{\bf r}|}.
\end{equation}
Here, the charge of the $I$-th nucleus is $Z_Ie$. 
The potential term for the Hamiltonian of electrons 
is given by the next operator. 
\begin{equation}
\hat{V}_{\rm ext} = \int d^3r v_{\rm ext}({\bf r}) \hat{n} ({\bf r}). 
\end{equation}
In the following discussion, we omit a constant Coulomb energy 
coming from the ion-ion interaction, 
\begin{equation}
V_{\rm ii}=\sum_{\langle I,J\rangle} \frac{Z_IZ_Je^2}{|{\bf R}_I-{\bf R}_J|},
\end{equation}
although this term is important for the charge neutrality condition. 
Hereafter, the symbol $\langle I,J\rangle$ denotes 
a pair of two different integers ranging from 1 to a finite 
integer (here it is $M$), and the summation with respect to these pairs 
are written as $\sum_{\langle I,J\rangle}$. 

If we have more than two electrons, 
the electron-electron interaction always takes place. 
In a static state, 
the inter-electron interaction is described by the next operator. 
\begin{eqnarray}
\label{Coulomb}
\hat{V}_{\rm ee} 
&=& 
\frac{1}{2} \int \! d^3r \, d^3r' \,
\frac{e^2}{|{\bf r}-{\bf r}'|} \sum_{\sigma,\sigma'}
\psi^\dagger_{\sigma}({\bf r}) \psi^\dagger_{\sigma'}({\bf r}') 
\psi_{\sigma'}({\bf r}') \psi_{\sigma}({\bf r}) \nonumber \\
&=& 
\frac{1}{2} \int \! d^3r \, d^3r' \,
\frac{e^2}{|{\bf r}-{\bf r}'|} 
: \hat{n}({\bf r}) \hat{n}({\bf r}') :
.
\end{eqnarray}
The symbol $:O:$ for an operator $O$ denotes the normal ordering of 
the field operators. In $:O:$, the order of $\psi$ and $\psi^\dagger$ 
is reorganized so that the creation operators come to the left 
of the annihilation operators by interchanging operators. 
If interchange of two field operators is made $m$ times, 
the sign of $(-1)^m$ is multiplied to the reordered operator. 

We consider an isolated electron system 
having $N\ge 2$ in an equilibrium. 
Competition between 
the Coulomb interaction $\hat{V}_{\rm ee}$ 
and other single-particle parts, 
$\hat{T}+\hat{V}_{\rm ext}$, causes various 
phase transitions in the electron systems. 
We search for the minimum energy allowed for the electron system. 
Thus any state vector, which has 
an anti-symmetric property for the spin-1/2 Fermion system, 
has to be normalizable, and has to have a finite kinetic energy. 
The class of wavefunctions for the allowed state vectors 
is $H^1({\bm R}^{3N})$. 

For any state vector $|\Psi\rangle$ of an $N$ electron state with $N\ge 2$, 
we have $|\Psi'\rangle \equiv 
\psi_{\sigma'}({\bf r}') \psi_{\sigma}({\bf r}) |\Psi\rangle$, 
and the quantum state $|\Psi'\rangle$ has a positive semidefinite norm, 
$\langle \Psi | : \hat{n}({\bf r}) \hat{n}({\bf r}') : 
| \Psi \rangle = \langle \Psi' | \Psi'\rangle \ge 0$. 
If we assume that $\langle \Psi|\hat{V}_{\rm ee}|\Psi \rangle = 0$, 
positivity of the Coulomb kernel requires that 
$\forall {\bf r}$ and $\forall {\bf r}'$, 
$\langle \Psi | : \hat{n}({\bf r}) \hat{n}({\bf r}') : | \Psi \rangle = 0$. 
But, then we have zero of the integral 
\[\int d^3rd^3r' 
\langle \Psi | : \hat{n}({\bf r}) \hat{n}({\bf r}') : | \Psi \rangle 
= \frac{N(N-1)}{2}= 0, \]
which yields $N=0$ or $N=1$. 

The expectation value of $\hat{V}_{\rm ee}$ by 
$\Psi\in H^1({\bm R}^{3N})$ is known to be finite.\cite{Lieb83} 
Actually, 
\begin{eqnarray}
\langle \Psi|\hat{V}_{\rm ee}|\Psi \rangle 
&=& 
\sum_{\langle i,j\rangle }
\int \prod_{l=1}^N d{\bf r}_l 
\frac{e^2}{|{\bf r}_i-{\bf r}_j|}
\Psi^*( \left\{{\bf r}_k,\sigma_k\right\})
\Psi(\left\{{\bf r}_{k'},\sigma_{k'}\right\}) 
\nonumber \\
&=& 
\sum_{\langle i,j\rangle }
\int \prod_{l=1}^N d{\bf r}_l 
\frac{e^2\theta(|{\bf r}_{ij}|)+e^2[1-\theta(|{\bf r}_{ij}|)]}{|{\bf r}_{ij}|}
\left|\Psi( \left\{{\bf r}_k,\sigma_k\right\})\right|^2
\nonumber \\
&\le& 
\sum_{\langle i,j\rangle }
\left\{
\int \prod_{l=1}^N d{\bf r}_l 
\frac{e^2\theta(|{\bf r}_{ij}|)}{|{\bf r}_{ij}|}
\int \prod_{l=1}^N d{\bf r}_l 
\left|\Psi( \left\{{\bf r}_k,\sigma_k\right\})\right|^2
\right.
\nonumber \\
&+&
\left.
\int \prod_{l=1}^N d{\bf r}_l 
e^2[1-\theta(|{\bf r}_{ij}|)]
\left|\Psi( \left\{{\bf r}_k,\sigma_k\right\})\right|^2 
\right\}
< \infty.
\end{eqnarray}
Here, we wrote ${\bf r}_i-{\bf r}_j={\bf r}_{ij}$ and 
$\theta(x)=1$ if $|x|\le 1$, $\theta(x)=0$ if $|x|>1$. 
Thus, for $N\ge 2$, we have,
\begin{equation}
0<\langle \Psi|\hat{V}_{\rm ee}|\Psi \rangle <\infty.
\end{equation}

\section{The charge density being the primary order parameter}
\label{Order_parameter}

The electron system is characterized by the electron charge density 
$n_\Psi({\bf r}) \equiv \langle \Psi | \hat{n}({\bf r}) | \Psi \rangle$. 
This physical quantity is observable by X-ray diffraction measurements. 
Since the wave function $\Psi(\left\{{\bf r}_i,\sigma_i\right\})$ 
is given as a function in the space $L^2$ of integrable functions, 
and since we assume that the kinetic energy is finite for $\Psi$, 
$n_\Psi({\bf r})^{1/2}$ and $\nabla n_\Psi({\bf r})^{1/2}$ is in $L^2$. 
Then, $n_\Psi({\bf r})$ is known to be also in $L^3$.\cite{Lieb83} 
The space $L^{3/2}+L^\infty$ is a dual of the $L^1+L^3$ space. 
Since the Coulomb potential is in $L^{3/2}+L^\infty$,
$v_{\rm ext}({\bf r})n_\Psi({\bf r})$ is integrable.  

The Hamiltonian of this electron system is given as 
\begin{equation}
\hat{H}=\hat{T}+\hat{V}_{\rm ee}+\hat{V}_{\rm ext}. 
\end{equation}
The total energy $E[\Psi,v_{\rm ext}]$ of this electron system is given by, 
\begin{equation}
E[\Psi,v_{\rm ext}] \equiv 
\langle \Psi | \hat{H} | \Psi \rangle
=
\langle \Psi | \hat{T} + \hat{V}_{\rm ee} |\Psi \rangle 
+ \int d^3r v_{\rm ext}({\bf r}) n_\Psi ({\bf r}). 
\end{equation}
The lowest steady state is given by minimizing this energy 
in the space of wavefunctions. 
\begin{equation}
E_0[v_{\rm ext}] = \min_{\Psi} E[\Psi,v_{\rm ext}]
= 
\min_{\Psi} \left[
\langle \Psi | \hat{T} + \hat{V}_{\rm ee} |\Psi \rangle 
+ \int d^3r v_{\rm ext}({\bf r}) n_\Psi ({\bf r}) \right]. 
\end{equation}

A definition of the order parameter is given by 
a derivative of the total energy with respect to the external field. 
For the electron system, this external degrees of freedom is 
given by $v_{\rm ext}({\bf r})$. 
Let's consider a stable electronic state $|\Psi\rangle$ without degeneracy. 
This state has an energy $E_0[v_{\rm ext}] =\langle\Psi|\hat{H}|\Psi\rangle$.  
If an infinitesimal variation of $v_{\rm ext}({\bf r})$ 
is given in $L^{3/2}+L^\infty$, the wavefunction minimizing 
$E[\Psi,v_{\rm ext}+\delta v_{\rm ext}]$ exists. 
This minimizing wavefunction may be given as 
\[|\Psi+\delta\Psi\rangle = e^{i\theta}\left(|\Psi\rangle +|\delta \Psi\rangle\right). \]
A complex phase factor $e^{i\theta}$ represents a gauge degree of freedom. 
We also use a notation, 
\begin{equation}
\delta \hat{H} = \int d^3r \delta v_{\rm ext}({\bf r}) \hat{n} ({\bf r}). 
\end{equation}
When $||\delta v_{\rm ext}|| = \delta \rightarrow 0$, 
the norm of $|\delta \Psi\rangle$ goes to zero, and 
the state $|\delta \Psi\rangle$ is orthogonal to $|\Psi\rangle$. 
Then, we have a derivative of $E_0[v_{\rm ext}]$. 
\begin{eqnarray}
\lefteqn{E_0[v_{\rm ext}+\delta v_{\rm ext}]
-E_0[v_{\rm ext}]}\nonumber \\
&=&
\frac{\langle \Psi | \hat{H}+\delta\hat{H} | \Psi + \delta \Psi \rangle}
{\langle \Psi | \Psi + \delta \Psi \rangle}
-
\frac{\langle \Psi + \delta \Psi | \hat{H} | \Psi \rangle}
{\langle \Psi + \delta \Psi | \Psi \rangle}
=
\frac{\langle \Psi + \delta \Psi | \delta\hat{H} | \Psi \rangle}
{\langle \Psi + \delta \Psi | \Psi \rangle}
\nonumber \\
&=&
\left(
\langle \Psi | \delta\hat{H} | \Psi \rangle
+\langle \delta \Psi | \delta\hat{H} | \Psi \rangle
\right)
\left(
1 - \langle \delta \Psi | \Psi \rangle + O(\delta^2) 
\right)
= \langle \Psi | \delta\hat{H} | \Psi \rangle + O(\delta^2) 
\nonumber \\
&=&
\int d^3r \delta v_{\rm ext}({\bf r}) n_\Psi ({\bf r})
=
\int d^3r \frac{\delta E_0[v_{\rm ext}]}{\delta v_{\rm ext}({\bf r})} 
\delta v_{\rm ext}({\bf r}) . 
\end{eqnarray}
This derivation follows a proof of the force theorem by Parr.\cite{Parr} 

In a steady state, since the state has to be stable against any perturbation, 
the functional derivative should exist. 
The coefficients of the functional derivative 
$\displaystyle
\frac{\delta E_0[v_{\rm ext}]}{\delta v_{\rm ext}({\rm r})}=n_\Psi ({\bf r})$ 
is the primary order parameter of the electron system. 
If we have a first-order phase transition by introducing 
$\delta v_{\rm ext}({\bf r})$, 
a level crossing in 
the lowest energy state $|\Psi\rangle$ occurs. 
Then, a jump in directed derivatives could be found. 
In such a special point, the derivative becomes ill-defined. 
However, this jump is detected as a jump in the charge density. 
In this sense, the electron charge density 
should be the primary order parameter. 

In section \ref{lambda_modified}, 
we will define another 
phase transition without a jump in $n({\bf r})$. 
This transition can happen in electron systems 
owing to internal degrees of freedom, {\it e.g.} the electron spin. 
The transition without change in $n({\bf r})$ 
can happen much frequently, 
when the interaction strength or the form of 
the inter-particle interaction are modified. 
Once the interaction strength is shifted from that of 
the Coulomb interaction, the resulting Hamiltonian describes 
a model system. The model can be identical to 
the Kohn-Sham model.\cite{Kohn-Sham} 
Thus, for the analysis of the Kohn-Sham scheme, 
it is important to analyze this second-type phase transitions 
in an abstract model space. 

\section{The density functional theory as a Landau theory}
\label{Landau_theory}

To define the universal energy density functional, 
a Sobolev space ${\cal I}_N$ of functions in ${\mathbb R}^3$ was 
introduced in the density functional theory. 
For a function $n({\bf r})$ in the set ${\cal I}_N$, 
$n({\bf r})\ge 0$ and $n^{1/2}({\bf r})$ 
is square integrable. Its gradient $\nabla n^{1/2}({\bf r})$ is also 
square integrable. In addition, $n({\bf r})$ satisfies, 
\begin{equation}\int d^3r n({\bf r}) = N. \end{equation}
We use the theorem 3.3 of Ref. \cite{Lieb83} 
stating that, if $n({\bf r})\in {\cal I}_N$, 
$F[n]$ given by the next definition exists. 
\begin{equation}
F[n]=\min_{\Psi\rightarrow n}
\langle \Psi | \hat{T}+\hat{V}_{\rm ee} | \Psi \rangle .
\end{equation}
The symbol $\Psi \rightarrow n$ represents that 
a minimizing state $|\Psi\rangle$ is searched 
with a constraint $\langle\Psi|\hat{n}({\bf r})|\Psi\rangle = n({\bf r})$. 
Then, the 
existence of the minimizing $\Psi$ is relevant. 

We have the next constrained minimization process. 
\begin{equation}
E_0[v_{\rm ext}] = \min_{n}
\min_{\Psi\rightarrow n} E[\Psi,v_{\rm ext}]
= 
\min_{n} \left[
F[n] + \int d^3r v_{\rm ext}({\bf r}) n ({\bf r}) \right]. 
\end{equation}
Here, $F[n]$ behaves as a free energy of the electron system. 

If we have a well-defined description of the free energy of 
the system as a functional of the primary order parameter, 
we have a complete expression of the Landau free energy. 
The density functional theory actually gives an example. 
In an $N$-representable form of the energy density functional, 
the Landau free energy of the system is given as a summation of 
the so-called universal energy density functional and 
the energy of the external scalar potential. 
The latter, $\displaystyle \int d^3r v_{\rm ext}({\bf r}) n_\Psi ({\bf r})$, 
is linear in the order parameter. 

\section{The $\lambda$ modified functional}
\label{lambda_modified}

Let $\lambda$ be a real parameter in $[0,1]$. 
We now consider a reduced energy density functional 
$F_\lambda[n]$ defined by,
\begin{equation}
F_\lambda [n]=\min_{\Psi'\rightarrow n}
\langle \Psi' | \hat{T}+\lambda \hat{V}_{\rm ee} | \Psi' \rangle .
\end{equation}
Existence of the minimizing $\Psi'$ is given also by the theorem 3.3 
of Ref. \cite{Lieb83}. 
We would like to address a next statement. 

\begin{Thm}
If $0\le \lambda \le 1$, and if $n({\bf r})\in {\cal I}_N$, 
$F_{\lambda}[n]$ is a monotone increasing continuous function of $\lambda$. 
$F_{\lambda}[n]$ is concave as a function of $\lambda$. 
\end{Thm}

First, choose $0\le \lambda_1<\lambda_2\le 1$. 
Assume that 
$F_{\lambda_1}[n] \ge F_{\lambda_2}[n]$. 
Choose a minimizing state 
$|\Psi_0\rangle \rightarrow n$ of the expectation value 
$\langle \Psi' | \hat{T}+\lambda_2 \hat{V}_{\rm ee} | \Psi' \rangle $. 
Then, we have,
\begin{eqnarray}
F_{\lambda_1}[n] 
&=& 
\min_{\Psi'\rightarrow n}
\langle \Psi' | \hat{T}+\lambda_1 \hat{V}_{\rm ee} | \Psi' \rangle 
\nonumber \\
&\ge&
\min_{\Psi'\rightarrow n}
\langle \Psi' | \hat{T}+\lambda_2 \hat{V}_{\rm ee} | \Psi' \rangle 
=
\langle \Psi_0 | \hat{T}+\lambda_2 \hat{V}_{\rm ee} | \Psi_0 \rangle 
\nonumber \\ 
&=&
\langle \Psi_0 | \hat{T}+\lambda_1 \hat{V}_{\rm ee} | \Psi_0 \rangle 
+(\lambda_2-\lambda_1)\times 
\langle \Psi_0 | \hat{V}_{\rm ee} | \Psi_0 \rangle 
\nonumber \\ 
&>&
\langle \Psi_0 | \hat{T}+\lambda_1 \hat{V}_{\rm ee} | \Psi_0 \rangle. 
\nonumber 
\end{eqnarray}
This inequality contradicts the definition of $F_{\lambda_1}[n]$. 
Thus, $F_{\lambda}[n]$ is a monotone increasing function of $\lambda$. 
Next, assume that $F_{\lambda}[n]$ is not continuous 
when $\lambda = \lambda_0 \ge 0$. 
This is equivalent to a statement 
that $\exists \varepsilon >0$, $\forall \delta >0$, 
$\exists \lambda > 0$, {\it s.t.} 
$\left\{
|\lambda - \lambda_0| < \delta \; \& \;
|F_{\lambda}[n] - F_{\lambda_0}[n]| > \varepsilon. 
\right\}$
For simplicity, let us further assume that 
$0<\lambda - \lambda_0 < \delta $ and then 
$F_{\lambda}[n] - F_{\lambda_0}[n]>\varepsilon$. 
If $n({\bf r})\in {\cal I}_N$, 
for any minimizing state vector $|\Psi_0\rangle$ of $F_{\lambda_0}[n]$, 
the wave function of $\Psi_0 \rightarrow n({\bf r})$ is in $H^1$, and 
\begin{equation}
0< \langle \Psi_0|\hat{V}_{\rm ee}|\Psi_0 \rangle <\exists C_0< \infty.
\end{equation}
If we let $\delta = \varepsilon/(2C_0)$, we have, 
\begin{eqnarray}
\lefteqn{
\min_{\Psi'\rightarrow n}
\langle \Psi' | \hat{T}+\lambda \hat{V}_{\rm ee} | \Psi' \rangle 
} \nonumber \\
&>& 
\min_{\Psi'\rightarrow n}
\langle \Psi' | \hat{T}+\lambda_0 \hat{V}_{\rm ee} | \Psi' \rangle + \varepsilon 
\nonumber \\
&=& 
\langle \Psi_0 | \hat{T}+\lambda_0 \hat{V}_{\rm ee} | \Psi_0 \rangle + \varepsilon 
\nonumber \\
&\ge& 
\langle \Psi_0 | \hat{T}+\lambda \hat{V}_{\rm ee} | \Psi_0 \rangle 
+(\lambda_0-\lambda) C_0 + \varepsilon 
\nonumber \\
&>&
\langle \Psi_0 | \hat{T}+\lambda \hat{V}_{\rm ee} | \Psi_0 \rangle 
-\delta C_0 + \varepsilon 
\nonumber \\
&=&
\langle \Psi_0 | \hat{T}+\lambda \hat{V}_{\rm ee} | \Psi_0 \rangle 
+\frac{\varepsilon}{2}\, .
\end{eqnarray}
This inequality contradicts the definition of the minimum. 

Consider 
$0\le \lambda_1 < \lambda_2 \le 1$, $0<\xi<1$, and 
$\lambda_\xi \equiv \xi\lambda_1+(1-\xi)\lambda_2$. 
We call a state 
$\Psi_{\lambda_\xi}\rightarrow n({\bf r})$, which minimizes 
$\langle \Psi' | \hat{T}+\lambda_\xi \hat{V}_{\rm ee} | \Psi' \rangle$. 
Then, we have 
\begin{eqnarray}
\lefteqn{F_{\xi\lambda_1+(1-\xi)\lambda_2}[n] } \nonumber \\
&=&
\xi
\langle\Psi_{\lambda_\xi}|\hat{T}+\lambda_1\hat{V}_{\rm ee}|\Psi_{\lambda_\xi}\rangle 
+(1-\xi)
\langle\Psi_{\lambda_\xi}|\hat{T}+\lambda_2\hat{V}_{\rm ee}|\Psi_{\lambda_\xi}\rangle 
\nonumber \\
&\ge& \xi F_{\lambda_1}[n]+(1-\xi) F_{\lambda_2}[n]. 
\label{Concavity}
\end{eqnarray}
This inequality ensures concavity of $F_{\lambda}[n]$ as a function of 
$\lambda$. $\blacksquare$

Following knowledge on 
the monotone increasing continuous functions and 
the convex (concave) functions, 
we immediately obtain results on derivatives 
and an integral of the derivative.\cite{Convex_analysis} 
Let's define Dini's derivatives of $F_\lambda[n]$, 
\begin{eqnarray}
D^-(\lambda )
&=&\lim_{y\rightarrow 0-} \sup_{y<h<0} \frac{F_{\lambda+h}[n]-F_\lambda[n]}{h},
\\
D_+(\lambda )
&=&\lim_{y\rightarrow 0+} \inf_{0<h<y} \frac{F_{\lambda+h}[n]-F_\lambda[n]}{h}.
\end{eqnarray}
When $\lambda=1$, we formally define $D_+(\lambda )$ 
by considering $F_\lambda[n]$ for $\lambda>1$. 
In practical simulation, however, $D_+(\lambda )$ is not required 
at $\lambda=1$. 

\begin{Cor}
For $n({\bf r})\in {\cal I}_N$, 
the monotone increasing concave function $F_\lambda[n]$ of $\lambda$ 
has a directed derivative 
at $\forall \lambda \in (0,1]$ and $D^-(\lambda ) \ge D_+(\lambda )$
\label{Cor3}
\end{Cor}

When $D^-(\lambda )=D_+(\lambda )$, 
we have the derivative, 
$\displaystyle \frac{d}{d\lambda} F_{\lambda}[n]$, 
of the monotone increasing function of $F_\lambda[n]$. 
The differentiability of a monotone increasing function 
is given by H. Lebesgue. Besides, the next statement holds. 

\begin{Cor}
If $0\le \lambda \le 1$, 
for $n({\bf r})\in {\cal I}_N$, 
$F_{\lambda}[n]$ is differentiable {\it a.e.} 
The derivative 
$\displaystyle \frac{d}{d\lambda} F_{\lambda}[n]$ is 
integrable in $[0,1]$. 
\label{Cor4}
\end{Cor}

Next, consider the case with $N\ge 2$. 
A set of state vectors 
$|\Psi\rangle$ reproducing $n({\bf r})$ is denoted as ${\cal Q}_n$. 
Fix $\lambda \in (0,1]$. 
We consider a subset of $|\Psi\rangle \in {\cal Q}_n$ minimizing 
$\langle \Psi | \hat{T}+\lambda \hat{V}_{\rm ee} 
| \Psi \rangle$ and call it ${\cal P}_n(\lambda)$. 
For any $|\Psi\rangle \in {\cal P}_n(\lambda)$, 
$0<\langle\Psi|\hat{V}_{ee}|\Psi\rangle<\infty$. 
Thus we have a finite range including values of 
$\langle\Psi|\hat{V}_{ee}|\Psi\rangle$ for 
$|\Psi\rangle \in {\cal P}_n(\lambda)$. 
The maximum of this range is given by a state vector 
in ${\cal P}_n(\lambda)$, and also the minimum is given by another vector. 
They might be different with each other. 
Thus they are denoted as 
$|\Psi_\lambda^-\rangle \in {\cal P}_n(\lambda)$ and 
$|\Psi_\lambda^+\rangle \in {\cal P}_n(\lambda)$. 
The very definitions of ${\cal P}_n(\lambda)$ 
and these vectors ensure that 
\begin{equation}
F_{\lambda}[n]
=\langle \Psi_{\lambda}^- | \hat{T}+\lambda \hat{V}_{\rm ee} 
| \Psi_{\lambda}^- \rangle 
=\langle \Psi_{\lambda}^+ | \hat{T}+\lambda \hat{V}_{\rm ee} 
| \Psi_{\lambda}^+ \rangle,
\label{flambda_eq}
\end{equation}
and that 
\begin{equation}
\langle\Psi_{\lambda}^+|\hat{V}_{ee}|\Psi_{\lambda}^+\rangle
\le \langle\Psi_{\lambda}^-|\hat{V}_{ee}|\Psi_{\lambda}^-\rangle.
\label{inequality-1}
\end{equation}

Now we prove existence of a vector giving 
the directed derivative of $F_\lambda[n]$. 
\begin{Lem}
For $\lambda\in (0,1]$, we have 
$D_+(\lambda )=\langle\Psi_{\lambda}^+|\hat{V}_{ee}|\Psi_{\lambda}^+\rangle$ 
and 
$D^-(\lambda )=\langle\Psi_{\lambda}^-|\hat{V}_{ee}|\Psi_{\lambda}^-\rangle$. 
\label{Lem4}
\end{Lem}

For $0 < \forall \delta < \lambda$, we have 
$0 < \lambda' =\lambda -\delta < \lambda$ and 
\begin{eqnarray}
F_{\lambda'}[n]
&\le&
\langle \Psi_{\lambda}^- | \hat{T}+\lambda' \hat{V}_{\rm ee} 
| \Psi_{\lambda}^- \rangle 
=
\langle \Psi_{\lambda}^- | \hat{T}+\lambda \hat{V}_{\rm ee} 
| \Psi_{\lambda}^- \rangle 
-\delta 
\langle\Psi_{\lambda}^-|\hat{V}_{ee}|\Psi_{\lambda}^-\rangle
\nonumber \\
&\le& 
\langle \Psi_{\lambda}^+ | \hat{T}+\lambda \hat{V}_{\rm ee} 
| \Psi_{\lambda}^+ \rangle 
-\delta 
\langle\Psi_{\lambda}^+|\hat{V}_{ee}|\Psi_{\lambda}^+\rangle
=
\langle \Psi_{\lambda}^+ | \hat{T}+\lambda' \hat{V}_{\rm ee} 
| \Psi_{\lambda}^+ \rangle 
.\end{eqnarray}
Thus, 
\begin{eqnarray}
D^-(\lambda )
&=&
\lim_{y\rightarrow 0+} \sup_{0<h<y} \frac{F_\lambda[n]-F_{\lambda-h}[n]}{h}
\nonumber \\
&\ge&
\lim_{y\rightarrow 0+} \sup_{0<h<y} 
\frac{F_\lambda[n]
-\langle \Psi_{\lambda}^- | \hat{T}+\lambda \hat{V}_{\rm ee} 
| \Psi_{\lambda}^- \rangle 
+h \langle\Psi_{\lambda}^-|\hat{V}_{ee}|\Psi_{\lambda}^-\rangle
}{h}
\nonumber \\
&=&\langle\Psi_{\lambda}^-|\hat{V}_{ee}|\Psi_{\lambda}^-\rangle.
\label{lower_ineq}
\end{eqnarray}
Similarly, we have, 
\begin{eqnarray}
D_+(\lambda )
&\le&
\langle\Psi_{\lambda}^+|\hat{V}_{ee}|\Psi_{\lambda}^+\rangle.
\label{upper_ineq}
\end{eqnarray}
These inequalities together with Eq.~(\ref{inequality-1})
give just another proof of Corollary \ref{Cor3}. 

Let's assume that $D^-(\lambda ) > 
\langle\Psi_{\lambda}^-|\hat{V}_{ee}|\Psi_{\lambda}^-\rangle$. 
Then, we have 
\begin{eqnarray}
0&<& 
\lim_{y\rightarrow 0+} \sup_{0<h<y} 
\frac{F_\lambda[n]-F_{\lambda-h}[n]
-h \langle\Psi_{\lambda}^-|\hat{V}_{ee}|\Psi_{\lambda}^-\rangle}{h}
\nonumber \\
&=&
\lim_{y\rightarrow 0+} \sup_{0<h<y} 
\frac{
\langle\Psi_{\lambda}^-|\hat{T}+(\lambda-h) 
\hat{V}_{ee}|\Psi_{\lambda}^-\rangle
-F_{\lambda-h}[n]}{h}.
\label{Ineq27}
\end{eqnarray}
Independently, we have a next inequality by the definition of 
$F_{\lambda-h}[n]$. 
\begin{equation}
\langle\Psi_{\lambda}^-|\hat{T}+(\lambda-h) \hat{V}_{ee}|\Psi_{\lambda}^-\rangle
\ge F_{\lambda-h}[n],
\end{equation}
which yields, $\forall h \in [0,y]$, 
\begin{equation}
f(h) \equiv 
\langle\Psi_{\lambda}^-|\hat{T}+(\lambda-h) \hat{V}_{ee}|\Psi_{\lambda}^-\rangle
- F_{\lambda-h}[n] \ge 0. 
\end{equation}
By Eq.~(\ref{flambda_eq}), $f(0)=0$. 

We show that $f(h)/h$ is also a monotone increasing function. 
Actually, if we assume that $0<\exists h_0 < y$, and 
that $\displaystyle \frac{f(h_0)}{h_0} > \frac{f(y)}{y}$, 
we have a next inequality. 
\[\left( 1- \frac{h_0}{y}\right) F_{\lambda} [n] 
+ \frac{h_0}{y} F_{\lambda -y}[n] > F_{\lambda - h_0}[n].\]
This contradicts to the concavity of $F_{\lambda}[n]$ 
given by Eq.~(\ref{Concavity}). 
Thus, Eq.~(\ref{Ineq27}) tells that 
\begin{equation}
0<
\lim_{y\rightarrow 0+} 
\frac{
\langle\Psi_{\lambda}^-|\hat{T}+(\lambda-y) 
\hat{V}_{ee}|\Psi_{\lambda}^-\rangle
-F_{\lambda-y}[n]}{y}.
\end{equation}
We call a state vector, which is in ${\cal P}_n(\lambda -y)$ 
and maximizes the expectation value of $\hat{V}_{ee}$, 
$|\Psi_{\lambda -y}^-\rangle$. 
Then the above expression yields, 
\begin{eqnarray}
\lim_{y\rightarrow 0+} 
\lefteqn{\langle \Psi_{\lambda -y}^-|\hat{V}_{ee}|\Psi_{\lambda -y}^-\rangle}
\nonumber \\
&>& 
\langle \Psi_{\lambda}^-|\hat{V}_{ee}|\Psi_{\lambda}^-\rangle
+\lim_{y\rightarrow 0+}
\frac{1}{y}\left[
\langle \Psi_{\lambda -y}^-|\hat{T}+\lambda\hat{V}_{ee}|\Psi_{\lambda -y}^-\rangle
-\langle \Psi_{\lambda}^-|\hat{T}+\lambda\hat{V}_{ee}|\Psi_{\lambda}^-\rangle
\right]
\nonumber \\
&\ge& 
\langle \Psi_{\lambda}^-|\hat{V}_{ee}|\Psi_{\lambda}^-\rangle. 
\label{Ineq31}
\end{eqnarray}
We note that $\lim_{y\rightarrow 0+}|\Psi_{\lambda -y}^-\rangle$ exists. 
The continuity of $F_\lambda[n]$ ensures that 
$\lim_{y\rightarrow 0+} |\Psi_{\lambda -y}^-\rangle \in {\cal P}_n(\lambda)$. 
The inequality, Eq.~(\ref{Ineq31}), 
contradicts to the definition of $|\Psi_{\lambda}^-\rangle$. 
Thus we conclude that 
$D^-(\lambda ) =
\langle\Psi_{\lambda}^-|\hat{V}_{ee}|\Psi_{\lambda}^-\rangle$. 
Similarly, 
$D_+(\lambda ) =
\langle\Psi_{\lambda}^+|\hat{V}_{ee}|\Psi_{\lambda}^+\rangle$. 
$\blacksquare$ 

For a fixed $n({\bf r})$, 
we now analyze the number of discontinuous points, where 
$D(\lambda) \equiv D^-(\lambda) - D_+(\lambda) > 0$.
Lemma \ref{Lem4} tells that, when $D(\lambda)>0$, 
${\cal P}_n(\lambda)$ has multiple elements 
which are distinguished by difference in 
the expectation value of $\hat{V}_{ee}$, 
and also in the expectation value of $\hat{T}$. 
At this discontinuous point, we have a change in 
the minimizing state from $|\Psi_{\lambda}^+\rangle$ 
to $|\Psi_{\lambda}^-\rangle$ by reducing $\lambda$. 
Therefore we have the next definition. 

\begin{Def}
A point where $D(\lambda)>0$ is called a transition point 
in the model space.
\end{Def}

We should note that {\it a priori} 
$|\Psi_\lambda^+\rangle$ and $|\Psi_\lambda^-\rangle$ 
are defined independently with each other. 
Lemma \ref{Lem4} implies that, 
for a point with $D(\lambda)=0$, we can identify 
$|\Psi_\lambda^+\rangle$ with $|\Psi_\lambda^-\rangle$ and 
replace one with the other. Thus 
the point with $D(\lambda)=0$ is not a level crossing point. 

\begin{Thm}
i) In $[0,1]$, we have at most a finite number of transition points, 
where $F_{\lambda}[n]$ has a finite discontinuity, which is 
greater than an arbitrary small number $\varepsilon$. 
ii) At an accumulation point of the discontinuous points 
$D(\lambda_i)$ ($i=1,2,\cdots,\infty$) with $\lambda_i \in (0,1]$, 
and $\lambda_i<\lambda_{i+1}$, we have 
$D(\lambda_i)\rightarrow 0$ for $i\rightarrow \infty$. 
\label{Theorem6}
\end{Thm}

i) For any finite number $\varepsilon>0$, 
we have a set $\Lambda(\varepsilon)$ of discontinuous points, 
at which $D(\lambda)>\varepsilon$. 
Assume that the number of elements of $\Lambda(\varepsilon)$ is 
more than the countable infinite. 
We can select a countable infinite subset of $\Lambda(\varepsilon)$ 
and name it $\bar{\Lambda}(\varepsilon)$. 
Points in $\bar{\Lambda}(\varepsilon)$ 
are in $[0,1]$ and are to be ordered as $\lambda_i<\lambda_{i+1}$. 
This is due to the selection axiom. 
Let's number the points in $\bar{\Lambda}(\varepsilon)$ 
as $\lambda_i$ ($i=1,2,\cdots$). 
We have, 
\[ D_+(\lambda_j) < D^-(\lambda_j) 
\le D_+(\lambda_{j-1}) < D^-(\lambda_{j-1}) ,\] 
\[D_+(\lambda_j) < D^-(\lambda_j)-\varepsilon 
< D^-(\lambda_{j-1})-2\varepsilon < D^-(\lambda_{1})-j\varepsilon .\] 
Since the set $\bar{\Lambda}(\varepsilon)$ is infinite, 
we have an integer $J>D^-(\lambda_1)/\varepsilon$ 
for which the derivative $D_+(\lambda_J)$ becomes strictly negative. 
This result contradicts to the increasing property of $F_\lambda[n]$. 
Thus the set $\Lambda(\varepsilon)$ has to be a finite set. 

ii) We consider $\inf_i D(\lambda_i)$ for $i=1, 2, \cdots$. 
Owing to i), a finite non-zero infimum is denied. 
Now consider $D=\lim_{N\rightarrow \infty} \sup_{i>N} D(\lambda_i)$. 
Assume that $D>0$. Then, 
we can find an integer $l_1$, for which $D(\lambda_{l_1})\ge D$. 
But, we can also find another integer $l_2>l_1$, 
for which $D(\lambda_{l_2})\ge D$, because of the definition of $D$. 
Then we have an infinite series of $\lambda_{l_i}$ ($i=1,2,\cdots$) 
with $D(\lambda_{l_i})\ge D>0$, which contradicts to i). 
Thus, $D=0$. 
$\blacksquare$ 

\begin{Lem}
For $n({\bf r}) \in {\cal I}_N$, 
the function 
$\displaystyle \frac{d}{d\lambda}F_\lambda[n]$ 
of $\lambda$ is Lipschitz continuous in $[0,1]$ 
and we have, 
\begin{equation}
\int_0^1 d\lambda \frac{d}{d\lambda} F_{\lambda}[n_{\Psi'}]
=F_{\lambda=1}[n]-F_{\lambda=0}[n]. 
\label{Integral}
\end{equation}
\label{Lem7}
\end{Lem}
For $\forall \Psi \in H^1$, we have a finite $R$ satisfying 
$\langle \Psi |\hat{H}_{\rm ee} |\Psi \rangle \le R$. 
Choose $R$ such that 
$D_+(0) \le R$. 
Since $F_\lambda[n]$ is continuous monotone increasing and concave, 
for $0 \le \forall \lambda_1 < \forall \lambda_2 \le 1$, 
we have, 
\begin{equation}
0\le F_{\lambda_2}[n] - F_{\lambda_1}[n] 
\le D_+(\lambda_1) (\lambda_2-\lambda_1)
\le D_+(0) (\lambda_2-\lambda_1)
\le R (\lambda_2-\lambda_1).
\end{equation}
Thus, we have Eq.~(\ref{Integral}). 
$\blacksquare$ 

\section{The Kohn-Sham minimization scheme}
\label{Kohn-Sham-minimization}

To discuss relevance of our theorems for the discussion of 
ground-state phase transitions, 
we introduce a Kohn-Sham minimization scheme.\cite{Hadjisavvas,Kusakabe-2001} 
For an external potential $v_{\rm ext}({\bf r})$, we 
have a ground state of the Coulomb system, 
$|\Psi_{\rm GS}\rangle$, which gives a ground-state density, 
$n_{\rm GS}({\bf r})$. First, we have a next equality. 
\begin{eqnarray}
\lefteqn{E_0[v_{\rm ext}]}\nonumber \\
 &=& 
\langle \Psi_{\rm GS}|\hat{T}+\hat{V}_{\rm ee}|\Psi_{\rm GS}\rangle 
+\int d^3r v_{\rm ext}({\bf r}) n_{\rm GS}({\bf r})
\nonumber \\
&=&
\min_{n}\left[F[n]+\int d^3r v_{\rm ext}({\bf r}) n({\bf r})\right]
\nonumber \\
&=&
\min_{n}\left[\min_{\Psi'\rightarrow n}
\langle \Psi'|\hat{T}|\Psi'\rangle
+F_{\lambda=1}[n]-F_{\lambda=0}[n]
+\int d^3r v_{\rm ext}({\bf r}) n({\bf r})\right]
\nonumber \\
&=&
\min_{n}\left[\min_{\Psi'\rightarrow n}\left\{
\langle \Psi'|\hat{T}|\Psi'\rangle
+\int_0^1 d\lambda \frac{d}{d\lambda} F_{\lambda}[n_{\Psi'}]
+\int d^3r v_{\rm ext}({\bf r}) n_{\Psi'}({\bf r})\right\}\right]
\nonumber \\
&=&
\min_{\Psi'}\left[
\langle \Psi'|\hat{T}|\Psi'\rangle
+\int_0^1 d\lambda \frac{d}{d\lambda} F_{\lambda}[n_{\Psi'}]
+\int d^3r v_{\rm ext}({\bf r}) n_{\Psi'}({\bf r})
\right]
\nonumber \\
&=&
\min_{\Psi'} G_{0,v_{\rm ext}}[\Psi'].
\end{eqnarray}
Here, we used a notation for the charge density given by 
$|\Psi\rangle$ as, 
\begin{equation}
n_{\Psi}({\bf r})\equiv \langle \Psi|\hat{n}({\bf r})|\Psi\rangle.
\end{equation}
We can show two conditions on the minimizing $\Psi$ 
of $G_{0,v_{\rm ext}}[\Psi]$. 
Note that, 
\begin{equation}
G_{0,v_{\rm ext}}[\Psi]
=
\langle \Psi|\hat{T}|\Psi\rangle
+F_{\lambda=1}[n_{\Psi}]-F_{\lambda=0}[n_{\Psi}]
+\int d^3r v_{\rm ext}({\bf r}) n_{\Psi}({\bf r}).
\end{equation}
Then we have, 
\begin{eqnarray}
G_{0,v_{\rm ext}}[\Psi]
&\ge&
\min_{\Psi'\rightarrow n_\Psi}\langle \Psi'|\hat{T}|\Psi'\rangle
+F_{\lambda=1}[n_{\Psi}]-F_{\lambda=0}[n_{\Psi}]
+\int d^3r v_{\rm ext}({\bf r}) n_{\Psi}({\bf r})
\nonumber \\
&=&
F_{\lambda=1}[n_{\Psi}]+\int d^3r v_{\rm ext}({\bf r}) n_{\Psi}({\bf r})
\nonumber \\
&\ge&
F_{\lambda=1}[n_{{\rm GS}}]
+\int d^3r v_{\rm ext}({\bf r}) n_{{\rm GS}}({\bf r}).
\label{G_0vP}
\end{eqnarray}
We see that equalities in Eq.~(\ref{G_0vP}) are satisfied, 
\begin{enumerate}
\item 
if $\Psi$ is identical to 
a states $\Psi'$, which minimizes 
$\langle \Psi'|\hat{T}|\Psi'\rangle$, 
with the constraint that $\Psi'\rightarrow n_\Psi$, 
\item 
and if $n_{\Psi}$ is identical to the true ground-state 
charge density $n_{\rm GS}({\bf r})$. 
\end{enumerate}
Thus, the minimizing process of $G_{0,v_{\rm ext}}[\Psi]$ 
gives us a state that reproduces $n_{\rm GS}({\bf r})$ and 
minimizes $\langle \Psi|\hat{T}|\Psi\rangle$. 
Two fundamental statements are addressed here. 
\begin{enumerate}
\item A minimizing state $\Psi$ of $G_{0,v_{\rm ext}}[\Psi]$ 
is the state which is searched in 
the constrained minimization of $F_{\lambda=0}[n_{\rm GS}]$. 
Thus $\Psi$ is the state motivated to be searched in the Kohn-Sham scheme. 
\item We do not need to have a secular equation to define 
the minimization process of $G_{0,v_{\rm ext}}[\Psi]$ for our discussion. 
\end{enumerate}

The definition of $G_{0,v_{\rm ext}}[\Psi]$ suggests us that 
we have a plenty of models for electron systems. 
Actually, we have another equality. 
\begin{eqnarray}
\label{variational_0_v}
E_0[v_{\rm ext}]
&=&
\min_{\Psi} G_{\lambda,v_{\rm ext}}[\Psi],
 \\
G_{\lambda,v_{\rm ext}}[\Psi]
&\equiv &
\langle \Psi|\hat{T}+\lambda V_{\rm ee}|\Psi\rangle
+\int_\lambda^1 d\lambda' \frac{d}{d\lambda'} F_{\lambda'}[n_{\Psi}]
+\int d^3r v_{\rm ext}({\bf r}) n_{\Psi}({\bf r})
.
\end{eqnarray}
The wave-function funcional, $G_{\lambda,v_{\rm ext}}[\Psi]$, 
determines $|\Psi_\lambda\rangle$, which may 
appear as a minimizing state in the definition of $F_\lambda[n_{\rm GS}]$. 
Therefore, when we fix $v_{\rm ext}({\bf r})$, minimization of 
$G_{\lambda,v_{\rm ext}}[\Psi]$ with $\lambda \in [0,1]$ 
produces a set of states $\left\{|\Psi_\lambda\rangle\right\}$ 
and thus the function $F_\lambda[n_{\rm GS}]$ of $\lambda$ as 
\begin{equation}
F_\lambda[n_{\rm GS}] = \langle\Psi_\lambda|\hat{T}+\lambda\hat{V}_{\rm ee}
|\Psi_\lambda\rangle .
\label{F_lambda_GS}
\end{equation}

We may introduce the Hartree term to formulate a much familiar 
form in the density functional theory. 
\begin{eqnarray}
\lefteqn{G_{0,v_{\rm ext}}[\Psi]} \nonumber \\
&=&
\langle \Psi|\hat{T}|\Psi\rangle
+\frac{1}{2} \int d^3r d^3r' \frac{e^2}{|{\bf r}-{\bf r}'|}
n_{\Psi}({\bf r})
n_{\Psi}({\bf r}')
+\int d^3r v_{\rm ext}({\bf r}) n_{\Psi}({\bf r})
\nonumber \\
&+&
\int_0^1 \left\{
d\lambda \frac{d}{d\lambda} 
\min_{\Psi'\rightarrow n_{\Psi}}
\langle \Psi|\hat{T}+\lambda \hat{V}_{\rm ee}|\Psi\rangle
-
\frac{1}{2}\int d^3r d^3r' \frac{e^2}{|{\bf r}-{\bf r}'|}
n_{\Psi}({\bf r})
n_{\Psi}({\bf r}')\right\}
\nonumber \\
&=&
\langle \Psi|\hat{T}|\Psi\rangle
+\frac{1}{2} \int d^3r d^3r' \frac{e^2}{|{\bf r}-{\bf r}'|}
n_{\Psi}({\bf r})
n_{\Psi}({\bf r}')
+\int d^3r v_{\rm ext}({\bf r}) n_{\Psi}({\bf r})
\nonumber \\
&+&
\int_0^1 
d\lambda 
\frac{1}{2}\int d^3r d^3r' \frac{e^2}{|{\bf r}-{\bf r}'|}
\left\{
\langle \Psi_\lambda|
:\hat{n}({\bf r})
\hat{n}({\bf r}'):
|\Psi_\lambda\rangle
-n_{\Psi}({\bf r})
n_{\Psi}({\bf r}')\right\}
\nonumber \\
&=&
\langle \Psi|\hat{T}|\Psi\rangle
+\frac{1}{2} \int d^3r d^3r' \frac{e^2}{|{\bf r}-{\bf r}'|}
n_{\Psi}({\bf r})
n_{\Psi}({\bf r}')
+\int d^3r v_{\rm ext}({\bf r}) n_{\Psi}({\bf r})
\nonumber \\
&+&
E_{\rm xc}[n_\Psi]. 
\end{eqnarray}
Similarly, we have 
\begin{eqnarray}
\lefteqn{G_{\lambda,v_{\rm ext}}[\Psi]} \nonumber \\
&=&
\langle \Psi|\hat{T}+\lambda\hat{V}_{\rm ee}|\Psi\rangle
+\frac{1-\lambda}{2} \int d^3r d^3r' \frac{e^2}{|{\bf r}-{\bf r}'|}
n_{\Psi}({\bf r})
n_{\Psi}({\bf r}')
+\int d^3r v_{\rm ext}({\bf r}) n_{\Psi}({\bf r})
\nonumber \\
&+&
\int_\lambda^1 
d\lambda' 
\frac{1}{2}\int d^3r d^3r' \frac{e^2}{|{\bf r}-{\bf r}'|}
\left\{
\langle \Psi_{\lambda'}|
:\hat{n}({\bf r})
\hat{n}({\bf r}'):
|\Psi_{\lambda'}\rangle
-n_{\Psi}({\bf r})
n_{\Psi}({\bf r}')\right\}
\nonumber \\
&=&
\langle \Psi|\hat{T}+\lambda\hat{V}_{\rm ee}|\Psi\rangle
+\frac{1-\lambda}{2} \int d^3r d^3r' \frac{e^2}{|{\bf r}-{\bf r}'|}
n_{\Psi}({\bf r})
n_{\Psi}({\bf r}')
+\int d^3r v_{\rm ext}({\bf r}) n_{\Psi}({\bf r})
\nonumber \\
&+&
E_{{\rm xc},\lambda}[n_\Psi]. 
\end{eqnarray}
We call $G_{\lambda,v_{\rm ext}}[\Psi]$ 
the $\lambda$-parametrized model energy functional. 
All of these models can determine $E_{0}[v_{\rm ext}]$ as shown by 
Eq.~(\ref{variational_0_v}) via a determination of $|\Psi_\lambda\rangle$. 
The charge density of $|\Psi_\lambda\rangle$ satisfies, 
$n_{\Psi_\lambda}({\bf r})=n_{\rm GS}({\bf r})$. 
When $0<\lambda<1$, 
since a reduced-interaction $\lambda \hat{V}_{\rm ee}$ 
appears in the definition of $G_{\lambda,v_{\rm ext}}[\Psi]$, 
its minimizing state, $|\Psi_\lambda\rangle$, is 
represented by a summation of Slater determinants. 

\section{Existence of $\varepsilon$ vicinity}
\label{epsilon-vicinity}

From now on, we consider possible level crossings 
in $F_\lambda[n_{\rm GS}]$ given in Eq.~(\ref{F_lambda_GS}). 
A level crossing occurs, when two or more states 
appear as minimizing states $|\Psi_\lambda\rangle$ 
of $\hat{T}+\lambda\hat{V}_{\rm ee}$ 
and when a jump in Dini's derivatives, $D(\lambda)>0$, happens. 

Existence of the crossing points is exemplified by 
a phase transition from 
the normal state to a ferromagnetic state recognized in 
the uniform electron gas system.\cite{Electron-gas,Ceperley_Alder} 
At the phase boundary, we have two uniform electron gas ground states, 
{\it i.e.} a paramagnetic state without spin polarization 
and a partially ferromagnetic state with a finite total spin. 
In general, we need to assume that crossing points 
appear at some of $\lambda \in [0,1]$. 
Once there appears a crossing point, 
we can count the number of crossing points 
on the $\lambda$ axis. 

First, we note that 
the level crossings happen at discretized points on the $\lambda$ axis. 
Two statements of Theorem \ref{Theorem6} deny 
a possibility to have crossing points with $D(\lambda)>0$ 
continuously or densely on the $\lambda$ axis. 
Next we need to consider a case with 
countable infinite numbers of crossing points 
in a finite region of $\lambda$. 
This case causes appearance of 
accumulation points of the crossing points on the $\lambda$ axis. 

Thus we can choose a finite interval of $\lambda$ satisfying 
one of two possible conditions for $D(\lambda)$: 
i) $\forall \lambda \in (\lambda_a,\lambda_b)$, $D(\lambda)=0$, 
or ii) one of the boundaries $\lambda_l$ ($l=a$ or $b$) 
is an accumulation point of $D(\lambda_i)>0$ and 
$\lim_{i\rightarrow \infty}D(\lambda_i)=0$. 
Our concern is whether the case ii) happens at $\lambda_b=1$ 
for a $v$-representable ground state density or not. 

Consider a unique ground state $|\Psi_{\rm GS}\rangle$ 
of an electron system in 
the external potential $v_{\rm ext}({\bf r})$, which 
has an electron charge density $n_{\rm GS}({\bf r})$. 
Simbolically, we write the $v$-representability of 
$n_{\rm GS}({\bf r})$ as $n_{\rm GS}({\bf r})\in{\cal A}_N$.\cite{Lieb83} 
The ground state is supposed to be stable against small perturbation. 
Let's assume that the point of $\lambda=1$ is 
an accumulation point of the level crossings 
along the $\lambda$ axis, and that we have the case ii). 

Crossing points are labeled as $\lambda_i$ ($i=1,2,\cdots,\infty$)
and $\lambda_i<\lambda_{i+1}$ for the present discussion. 
At any level crossing point $\lambda_i$, owing to Lemma \ref{Lem4}
there are at least two minimizing states, $|\Psi_{\lambda_i}^+\rangle$ 
and $|\Psi_{\lambda_i}^-\rangle$. 
The expectation values $\langle\hat{T}+\lambda_i\hat{V}_{\rm ee}\rangle$ 
by these states are the same at the crossing point. 
However, $|\Psi_{\lambda_i}^+\rangle$ 
and $|\Psi_{\lambda_i}^-\rangle$ are distinguished by 
$\langle \hat{V}_{ee} \rangle$ and $\langle \hat{T} \rangle$. 
Thus, they are linearly independent as state vectors. 
We can also show that 
$\langle\Psi_{\lambda_i}^+|\hat{V}_{\rm ee}|\Psi_{\lambda_i}^+\rangle$ 
decreases, when $i$ increases, as exemplified in the proof of 
Theorem \ref{Theorem6}. 

In between two neighboring crossing points, we have 
a continuous change in $|\Psi_\lambda\rangle$ and $D(\lambda)=0$. 
The state, $|\Psi_{\lambda_i}^+\rangle$ is connected to 
$|\Psi_{\lambda_{i+1}}^-\rangle$. 
We may select $|\Psi_{\lambda_i}^+\rangle$ as 
a representative state for this finite range $[\lambda_i,\lambda_{i+1}]$. 

Existence of a big number of level crossing points 
at a close vicinity of this accumulation point requires 
existence of a plenty number of nearly degenerate states, 
$|\Psi_{\lambda_i}^+\rangle$ with $i=1,2,\cdots,\infty$. 
Here, the state, $|\Psi_{\lambda_i}^+\rangle$, 
might not be an eigen state of any 
potential problem, but minimizes just the expectation value of 
$\langle\hat{T}+\lambda \hat{V}_{\rm ee}\rangle$ keeping the density. 
They are not distinguished by the primary order parameter, $n({\bf r})$, 
nor by difference in any external symmetry breaking 
observed by the external potential $v_{\rm ext}({\bf r})$. 
Only internal degrees of freedom distinguish these infinite numbers of 
$|\Psi_{\lambda_i}^+\rangle$. 

We should note that only to have these states 
does not directly mean existence of infinite numbers of 
degenerate eigen states at $\lambda=1$. 
This is because we have a possibility that 
all of $|\Psi_{\lambda_i}^+\rangle$ except for $|\Psi_{\rm GS}\rangle$ 
become non-eigen-states of $\hat{H}$, but are just variational states. 
Furthermore, if continuous change in $|\Psi_\lambda\rangle$ 
is allowed in a finite range of $\lambda$, 
only one minimizing state appears as a unique minimum 
for $\langle\hat{T}+\lambda\hat{V}_{\rm ee}\rangle$ in the range. 

However, the state $|\Psi_{\lambda_i}^+\rangle$ exists and 
may be used as variational states at any $\lambda\in[0,1]$. 
By a simple inspection, we can see that 
at $\lambda=1$, the variational energies of 
$\langle\Psi_{\lambda_i}^+|\hat{T}+\hat{V}_{\rm ee}|\Psi_{\lambda_i}^+\rangle$ 
are separated by a finite gap with each other. 
However, when $\lambda=1$ is an accumulation point, we have 
$|\langle\Psi_{\lambda_i}^+|\hat{T}
+\hat{V}_{\rm ee}|\Psi_{\lambda_i}^+\rangle
-\langle\Psi_{\lambda_{i+1}}^+|\hat{T}
+\hat{V}_{\rm ee}|\Psi_{\lambda_{i+1}}^+\rangle|\rightarrow 0$ 
for $i\rightarrow \infty$. 
Thus the gap for $i\gg N$ should be much smaller than 
any energy separation in the energy spectrum 
of the finite size system. 

To discuss the accumulation points further, 
we now analyze existence or non-existence 
of a potential $v_\lambda({\bf r})$ 
which gives a secular equation 
determining a state vector $|\Phi_\lambda\rangle$. 
Namely, for $\lambda =1-\delta\lambda$ with $0<\delta\lambda\ll 1$, 
we formulate a quantum mechanical problem, 
whose solution $|\Phi_\lambda\rangle$ satisfies 
$||\langle\Phi_\lambda|\hat{n}({\bf r})|\Phi_\lambda \rangle
-n_{\rm GS}({\bf r})||_\infty=0$, and 
$|\Phi_\lambda \rangle \rightarrow |\Psi_{\rm GS}\rangle$ 
for $\lambda\rightarrow 1$. 
Simbolically, we write this statement on 
another $v$-representability of $n_{\rm GS}({\bf r})$ 
as $n_{\rm GS}({\bf r})\in{\cal A}_{\lambda,N}$. 

For simplicity, we consider a compact space 
by introducing a cube with volume $L^3$ under 
the periodic boundary condition. 
Then, we can introduce a Fourier series expansion 
for the potential $v_\lambda({\bf r})$, 
which is used as the Lagrange multiplier to fix 
the charge density of $|\Phi\rangle$, as, 
\begin{equation}
v_\lambda({\bf r}) = 
\sum_{{\bf G}} v_{\lambda,{\bf G}} \exp (i{\bf G}\cdot{\bf r}). 
\end{equation}
Consider an $N$ electron state $|\Phi\rangle$. 
We introduce a functional $Q[|\Phi\rangle, 
v_{\lambda,{\bf G}}, E: \delta \lambda]$ as,
\begin{eqnarray}
\lefteqn{Q[|\Phi\rangle, 
v_{\lambda,{\bf G}}, E: \delta \lambda]}\nonumber \\
&=&
\langle\Phi|\hat{T}+(1-\delta \lambda) \hat{V}_{\rm ee} 
+ 
\int d^3r v_{\rm ext}({\bf r})\hat{n}({\bf r})
|\Phi\rangle \nonumber \\
&+&
\int d^3r v_{\lambda}({\bf r})
\left[\langle\Phi|\hat{n}({\bf r})|\Phi\rangle -n_{\rm GS}({\bf r})\right]
- E \left[\langle\Phi|\Phi\rangle - 1\right]
\nonumber \\
&=&
\langle\Phi|\hat{T}+\hat{V}_{\rm ee} 
+ \int d^3r v_{\rm ext}({\bf r})\hat{n}({\bf r})
|\Phi\rangle \nonumber \\
&+&
\langle\Phi|
\left\{
\sum_{{\bf G}\neq {\bf 0}}
\sum_{{\bf G}'}\sum_\sigma
v_{\lambda,{\bf G}}
c^\dagger_{{\bf G}'-{\bf G},\sigma} c_{{\bf G}',\sigma}
-\delta \lambda \hat{V}_{\rm ee} 
\right\}
|\Phi\rangle 
\nonumber \\
&-& 
\sum_{{\bf G}\neq {\bf 0}}
v_{\lambda,{\bf G}}
n_{\rm GS}(-{\bf G}) 
-\bar{E} \left[\langle\Phi|\Phi\rangle - 1\right].
\end{eqnarray}
Here, 
$n_{\rm GS}({\bf G})$ is the Fourier component of $n_{\rm GS}({\bf r})$, 
$c_{{\bf G},\sigma}$ are electron annihilation operators 
with the spin $\sigma$, and 
$\bar{E}=E-Nv_{\lambda,{\bf 0}}$. 
By making derivatives of $Q$ with respect to variables 
except for a parameter $\delta \lambda$, 
we have next secular equations. 
\begin{eqnarray}
\left\{\hat{H}+
\hat{H}_{\delta \lambda}
\right\}|\Phi\rangle = \bar{E} |\Phi\rangle, 
\label{Eig-1}
\\
\hat{H}_{\delta \lambda}
=
\sum_{{\bf G}\neq {\bf 0}}
\sum_{{\bf G}'}\sum_\sigma
v_{\lambda,{\bf G}}c^\dagger_{{\bf G}'-{\bf G},\sigma} c_{{\bf G}',\sigma}
-\delta \lambda \hat{V}_{\rm ee} 
\\
\langle\Phi|\Phi\rangle = 1,
\label{Eig-2}
\\
\sum_{{\bf G}'}\sum_\sigma
\langle\Phi|
c^\dagger_{{\bf G}'+{\bf G},\sigma} c_{{\bf G}',\sigma}
|\Phi\rangle 
=
n_{\rm GS}({\bf G}) .
\label{Eig-3}
\end{eqnarray}

At $\lambda=1$, $\delta\lambda=0$, 
the solution of Eqs.~(\ref{Eig-1}), (\ref{Eig-2}), and (\ref{Eig-3}) 
is given by the normalized state 
$|\Psi_{\rm GS}\rangle$ 
with $v_{\lambda,{\bf G}}=0$ for all ${\bf G}$. 
For any $v_{\lambda}({\bf r})\in L^{3/2}+L^\infty$, 
or equivalently for any set of $v_{\lambda,{\bf G}}$, 
with $\delta \lambda >0$, 
we have a normalized eigen state $|\Phi\rangle$ 
of Eqs.~(\ref{Eig-1}) and (\ref{Eig-2}). 
So the construction of $v_{\lambda,{\bf G}}$ 
to meet Eq.~(\ref{Eig-3}) is the problem. 
If $v_{\lambda,{\bf G}}$ exists, 
$n_{\rm GS}({\bf r})\in{\cal A}_{\lambda,N}$. 

\begin{Lem}
When a unique ground state of $\hat{H}$ gives 
the charge density $n_{\rm GS}({\bf r})$, 
and if there happens 
accumulation of points $\lambda_i$ ($i=1,2,\cdots,\infty$) 
satisfying $0<\lambda_i<\lambda_{i+1}<1$ and $D(\lambda_i)>0$, 
$n_{\rm GS}({\bf r})$ is not in ${\cal A}_{\lambda_i,N}$. 
We have a series of potentials 
$v_{\lambda_i}({\bf r})$, which gives 
a set of pure states $|\tilde{\Psi}_{\lambda_i}\rangle$, and 
$||\langle\tilde{\Phi}_{\lambda_i}
|\hat{n}({\bf r})|\tilde{\Phi}_{\lambda_i} \rangle
-n_{\rm GS}({\bf r})||_\infty \rightarrow 0$
as $i\rightarrow \infty$. 
\end{Lem}

To show this lemma, 
we classify possible conditions and deny possibilities 
except for a case that $n_{\rm GS}({\bf r})$ is not in 
${\cal A}_{\lambda_i,N}$, which is the case 4 below. 
\begin{description}
\item[case 1] 
There exists $v_{\lambda_i}({\bf r})$, and 
$|\Psi_{\lambda_i}^+\rangle$ and 
$|\Psi_{\lambda_i}^-\rangle$ are eigen states of 
$\displaystyle \hat{H}+\hat{H}_{\delta \lambda} 
=\hat{T}+\lambda_i \hat{V}_{ee}
+\int d^3r v_{\lambda_i}({\bf r})\hat{n}({\bf r})$.
\item[case 2] 
There exist 
$v_{\lambda_i}^+({\bf r})$ and 
$v_{\lambda_i}^-({\bf r})$, 
which are different from each other more than a constant. 
Two states, $|\Psi_{\lambda_i}^\pm\rangle$, 
are eigen states of 
$\displaystyle \hat{T}+\lambda \hat{V}_{ee}+\int d^3r 
v^\pm_{\lambda_i}({\bf r})\hat{n}({\bf r})$, respectively. 
\item[case 3] 
One of $|\Psi_{\lambda_i}^+\rangle$ and 
$|\Psi_{\lambda_i}^-\rangle$ is an eigen state of 
$\displaystyle \hat{T}+\lambda \hat{V}_{ee}
+\int d^3r v_{\lambda_i}({\bf r})\hat{n}({\bf r})$, 
and the other is not. 
\item[case 4] 
Both of minimizing states, 
$|\Psi_{\lambda_i}^+\rangle$ and 
$|\Psi_{\lambda_i}^-\rangle$, are not 
eigen states of any potential problem given as 
$\displaystyle \hat{T}+\lambda \hat{V}_{ee}
+\int d^3r v_{\lambda_i}({\bf r})\hat{n}({\bf r})$. 
\end{description}

We first deny the case 1. 
Let's assume that the solution 
$v_{\lambda,{\bf G}}$ exists for $n_{\rm GS}({\bf r})$. 
It means that we have a minimum of $Q$, which is given by 
an eigen state $|\Phi\rangle$ of Eq.~(\ref{Eig-1}). 
If we insert $|\Psi_\lambda\rangle$ in the functional $Q$, we have 
\begin{eqnarray}
\lefteqn{Q[|\Psi_\lambda\rangle, 
v_{\lambda,{\bf G}}, E: \delta \lambda]}
\nonumber \\
&=&
\langle \Psi_\lambda|\hat{T}+\lambda\hat{V}_{\rm ee}|\Psi_\lambda\rangle 
+\int d^3r v_{\rm ext}({\bf r}) n_{\rm GS}({\bf r})
\nonumber \\
&\le&
\langle \Phi|\hat{T}+\lambda\hat{V}_{\rm ee}|\Phi\rangle 
+\int d^3r v_{\rm ext}({\bf r}) n_{\rm GS}({\bf r})
\nonumber \\
&=&
Q[|\Phi\rangle, 
v_{\lambda,{\bf G}}, E: \delta \lambda]. 
\end{eqnarray}
Thus, by the variational principle of the quantum mechanics, we see that 
$|\Psi_\lambda\rangle$ is also an eigen state of Eq.~(\ref{Eig-1}). 
If we choose $\lambda=\lambda_i$, 
we have two degenerate eigen states, 
$|\Psi_{\lambda_i}^+\rangle$ and 
$|\Psi_{\lambda_i}^-\rangle$. 
They are orthogonal with each other. 
By taking the limit $i\rightarrow\infty$, 
we have two limiting states, 
$|\Psi_{\lambda_\infty}^+\rangle$ and 
$|\Psi_{\lambda_\infty}^-\rangle$, 
since these state vectors are in a Banach space 
and existence of the weak limit is ensured by 
$\langle \Psi_{\lambda_i}^\pm|\hat{V}_{\rm ee}|\Psi_{\lambda_i}^\pm\rangle 
\rightarrow 
\langle \Psi_{\rm GS}|\hat{V}_{\rm ee}|\Psi_{\rm GS}\rangle$ 
owing to Theorem \ref{Theorem6}. 
Orthogonality between 
$|\Psi_{\lambda_\infty}^+\rangle$ and 
$|\Psi_{\lambda_\infty}^-\rangle$ holds. 
Of course $\lambda_\infty=1$. 
These states minimize $Q[|\Psi\rangle, 
v_{\lambda=1,{\bf G}}, E: \delta \lambda=0]$. 
This means that the ground state has to be degenerate, 
which contradicts to the uniqueness of the ground state 
providing $n_{\rm GS}({\bf r})$. 

If we have a case 2, we have two independent potential 
$v_{\lambda_i}^+({\bf r})$ and 
$v_{\lambda_i}^-({\bf r})$, 
both of which give the same density $n_{\rm GS}({\bf r})$. 
This case contradicts the Hohenberg-Kohn theorem\cite{Hohenberg-Kohn} 
and thus it is denied at any $i$. 

If we have a case 3, we again have a difficulty. 
Both of $|\Psi_{\lambda_i}^+\rangle$ and 
$|\Psi_{\lambda_i}^-\rangle$ give the same variational energy for 
$\displaystyle \hat{T}+\lambda \hat{V}_{ee}
+\int d^3r v_{\lambda_i}({\bf r})\hat{n}({\bf r})$. 
The variational principle tells that 
a state having the variational energy of the lowest eigen state 
gives a degenerate eigen state. 
This fact contradicts an assumption that one of these two states 
is not an eigen state of any potential problem. 

So, we conclude the case 4, which says that 
$v_{\lambda_i,{\bf G}}$ does not exist, 
when $\lambda=1$ is an accumulation point. 
Since $n_{\rm GS}({\bf r}) \in {\cal I}_N$, 
we know existence of an $N$-particle density matrix $\Gamma$, 
which gives the infimum of the Lieb functional.\cite{Lieb83} 
$\Gamma$ has to be that for a mixed state, 
\begin{equation}
\Gamma = \sum_l C^{l}_{\lambda_i} 
|\Psi^{l}_{\lambda_i}\rangle \langle\Psi^{l}_{\lambda_i}|.
\end{equation}
$|\Psi^{l}_{\lambda_i}\rangle$ is given as 
an eigen state of a potential problem and 
these states are orthogonal with each other. 
Since the limit of $\lambda_i\rightarrow 1$ is given as 
a pure state, when $i\rightarrow \infty$, 
coefficients $C^{l}_{\lambda_i}$ and state vectors satisfy, 
\begin{eqnarray}
C^{l}_{\lambda_i} &\rightarrow& \delta_{l,l_0}, \\
|\Psi^{l_0}_{\lambda_i}\rangle &\rightarrow& |\Psi_{\rm GS}\rangle. 
\end{eqnarray}
Thus the second statement of the lemma holds. $\blacksquare$

The meaning of this lemma is somewhat redundant. 
When we have an accumulation point at $\lambda=1$, 
we have no potential series, $v_{\lambda_i}({\bf r})$, 
whose pure ground state reproduces exactly $n_{\rm GS}({\bf r})$. 
Thus the case 4, if it is found, 
gives an example of an $N$-representable density 
which is in ${\cal A}_N$ but not in ${\cal A}_{\lambda,N}$ for $\lambda<1$. 
This density is apparently not pure-state $v$-representable 
in any Kohn-Sham scheme. 
The accumulation of crossing points for a ground state density 
contradicts a picture of the ordinal Kohn-Sham scheme, 
which assumes that a unique ground state of the Coulomb problem is 
reproduced by a non-interacting system with an optimized potential. 

In this case, we should follow the present lemma to know 
existence of a converging series of quantum mechanical models. 
Actually, the potential series of $v_{\lambda_i}({\bf r})$, 
whose ground state density is slightly different from 
the final solution, 
can be used to find a Cauchy sequence of $n_{\lambda_i}({\bf r})$ 
converging to $n_{\rm GS}({\bf r})$. 
This converging series is found in the model space of 
the multi-reference generalization of the Kohn-Sham scheme. 
We may call the region of the model space as 
an $\varepsilon$-vicinity, in which 
a convergence of a simulation is guaranteed. 

Conversely, if a density $n_{\rm GS}({\bf r})$ is 
pure-state $v$-representable in a Kohn-Sham scheme 
including a multi-reference generalization, 
there is no accumulation of crossing points at $\lambda=1$. 
In this case, we have a well-defined $\varepsilon$-vicinity 
around the true ground state in the model space, 
in which no level crossing is found. 
More precisely, we have a next statement. 
If the density $n_{\rm GS}({\bf r})$ 
is $v$-representable also for $\lambda<1$, 
accumulation of phase transition points is forbidden 
when $\lambda\rightarrow 1$. 
Finding a convergence in the density searched in 
an optimization process of model quantum systems 
may allow us to conclude 
no remaining level-crossing point along a line approaching 
in the true Coulomb system given by $\hat{H}$. 
Therefore, we conclude the 
existence of an $\varepsilon$ vicinity 
around the Coulomb system in the model space, 
where no essential phase transition occurs in the direction 
to reduce the interaction strength by introducing $\lambda <1$, 
and keeping the charge density $n({\bf r})$ unchanged. 

Now we re-analyze the potential, $v_{\lambda,{\bf G}}$, 
in a pertabative argument. 
Although this problem was treated more elegantly by Kohn,\cite{Kohn} 
we want to show a subtle problem on existence of $v_{\lambda,{\bf G}}$. 
We have the limiting solution of $|\Psi_{\rm GS}\rangle$ 
with $v_{\lambda,{\bf G}}=0$, 
which is an eigen solution of Eq.~(\ref{Eig-1}) 
at least as a stationary state. 
The construction of $Q$ is based on 
the quantum mechanical variational principle, and 
the limit of $Q$ for $\lambda\rightarrow 1$ 
should behave regularly. 
In order to inspect on the $v$-representability of 
the normal solution $n_{\rm GS}({\bf r})$ for 
$\lambda<1$, we argue a perturbative construction method 
of $v_{\lambda,{\bf G}}$. 
When the spectrum of $\hat{H}$ is normal, 
and when the excited states are written as $|\Psi_i\rangle$, 
the solution of Eq.~(\ref{Eig-1}) is represented 
by the ordinal perturbation theory as, 
\begin{equation}
|\Phi\rangle 
=
\frac{1}{C}\left\{
|\Psi_{\rm GS}\rangle 
+
\sum_i
\frac{1}{E_{0}-E_i}
|\Psi_i\rangle
\langle\Psi_i| 
\hat{H}_{\delta \lambda}
|\Psi_{\rm GS}\rangle 
+o(\delta \lambda)\right\}.
\end{equation}
The expectation value in Eq.~(\ref{Eig-3}) gives, 
\begin{eqnarray}
\lefteqn{
\sum_{{\bf G}'}\sum_\sigma
\langle\Phi|
c^\dagger_{{\bf G}'+{\bf G},\sigma} c_{{\bf G}',\sigma}
|\Phi\rangle 
-
n_{\rm GS}({\bf G})}
\nonumber \\
&=&
\frac{1}{C^2}\left\{
\sum_{{\bf P}\neq {\bf 0}}
\sum_i
\frac{1}{E_{0}-E_i}
\langle \Psi_{\rm GS}|
\sum_{{\bf G}'}\sum_\sigma
c^\dagger_{{\bf G}'+{\bf G},\sigma} c_{{\bf G}',\sigma}
|\Psi_i\rangle \right.
\nonumber \\
&&\times
\langle\Psi_i| 
\sum_{{\bf G}''}\sum_\sigma
c^\dagger_{{\bf G}''-{\bf P},\sigma} c_{{\bf G}'',\sigma}
|\Psi_{\rm GS}\rangle 
v_{\lambda,{\bf P}}
\nonumber \\
&-&
\sum_i\left.
\frac{1}{E_{0}-E_i}
\langle \Psi_{\rm GS}|
\sum_{{\bf G}'}\sum_\sigma
c^\dagger_{{\bf G}'+{\bf G},\sigma} c_{{\bf G}',\sigma}
|\Psi_i\rangle 
\langle\Psi_i| 
\delta \lambda \hat{V}_{\rm ee} 
|\Psi_{\rm GS}\rangle 
+{\rm c.c.}+ o(\delta \lambda)
\right\}
\nonumber \\
&=&\sum_{{\bf P}\neq {\bf 0}}A_{{\bf G},{\bf P}}
v_{\lambda,{\bf P}}
+
\delta \lambda B_{{\bf G}} + {\rm c.c.} + o(\delta \lambda).
\label{Eig-3c}
\end{eqnarray}
Here, $C$ is a normalization constant. 
In general, we have a non-zero vector $B_{{\bf G}}$. 
Since the matrix $A_{{\bf G},{\bf P}}$ is an Hermite matrix, 
we have a solution 
$v_{\lambda,{\bf P}}$ of Eq.~(\ref{Eig-3}) 
in $o(\delta \lambda)$. 
Thus, we may utilize the determination method of 
$v_{\lambda,{\bf P}}$ to analyze the Kohn-Sham method, 
in which the density $n_{\rm GS}({\bf r})$ is reproduced by 
another artificial model system. 

However, we find a difficulty in Eq.~(\ref{Eig-3c}). 
The required conditions amount to the same number as 
$v_{\lambda,{\bf P}}$. 
So, once we consider higher-order conditions 
for $\lambda^n$ as independent, 
the number of the conditions is over the number of 
$v_{\lambda,{\bf P}}$. 
This suggests that the solution could be found in 
a non-local potential. 
Even if so, the number of conditions is too large. 
If we consider all the expansions in Eq.~(\ref{Eig-3c}) 
as functions of $\lambda$ and $v_{\lambda,{\bf P}}$, 
the determination equations form 
simultaneous equations of $v_{\lambda,{\bf P}}$. 
The structure is not trivial for higher order terms. 
Since they are non-linear determination equations, 
the solution is expected to exist 
only when all the expressions are derived exactly. 

Before closing this section, we summarize 
conditions for Lemma \ref{Lem4} and Theorem \ref{Theorem6}. 
The proofs of these statements tell that 
the expectation value of the relevant interaction term 
has to be positive and finite, 
{\it i.e.}, $0< \langle \hat{V}_{ee} \rangle<\infty$. 
The convexity (or precisely the concavity nature of $F_\lambda[n]$) 
comes from linear nature of the $\lambda$-modified $F_\lambda[n]$ 
with respect to the parameter $\lambda$. 
The constrained minimization allows us to conclude 
both continuous nature of the functional and 
existence of minimizing states, which give 
Dini's derivative of the parametrized energy density functional. 

\section{Summary and conclusions}
\label{Summary_conclusions}

In section \ref{lambda_modified}, 
we consider the $\lambda$ modification of $F[n]$, while 
the primary order parameter $n({\bf r})$ does not change. 
We consider a space of models, $G_{\lambda,v_{\rm ext}}[\Psi]$, 
with $\lambda \in [0,1]$ in section \ref{Kohn-Sham-minimization}. 
In the model space, at a discontinuous point of $F_\lambda[n]$ 
with fixed $n({\bf r})$, 
the minimizing states, $|\Psi_\lambda^+\rangle$ and 
$|\Psi_\lambda^-\rangle$, exist. 
Existence of discontinuous points may be detected by 
$\displaystyle \frac{d}{d\lambda}F_\lambda[n]$, or 
$\displaystyle \frac{d^2}{d\lambda^2}F_\lambda[n]$. 
The latter behaves as a delta function at the discontinuous points. 
We can use these functions as an indicator for 
discontinuous transition points in a set of states giving 
$\Psi\rightarrow n({\bf r})$. 

Owing to the above discussion, we have a physical conclusion on 
the universal energy density functional. 
Consider a true density $n_{\rm GS}({\bf r})$, 
which is given by a ground state 
in an external potential $v_{\rm ext}({\bf r})$. 
If the state is non degenerate (or at least finitely degenerated) 
and if it is stable, 
we have no accumulation point of the level crossing points 
for $F_\lambda[n_{\rm GS}]$ at $\lambda=1$ 
in the searching process utilizing a potential problem with 
the reduced interaction strength. 
Thus, when we move away from $\lambda=1$ on the $\lambda$ axis, 
we have an $\varepsilon$ vicinity, where no level crossing 
happens in the model space, which is to be defined by a secular equation 
of the many-Fermion system with the reduced interaction. 

In a multi-reference density functional theory (MR-DFT), 
one of the authors considered general modification of the universal energy 
density functional.\cite{Kusakabe-2001,Kusakabe-JPCM} 
In addition to the Kohn-Sham description of the true density,\cite{Kohn-Sham} 
we have plenty of effective descriptions using model Fermion systems. 
The models include partially correlated systems, whose 
ground state is obtained in a multiple Slater determinants. 
We can introduce distance between two models given by the 
norm of the charge density and then 
the set of models with the charge distance becomes a space of models. 
An advantage of this generalization of DFT is that 
we can search for an optimized model in a space of models including 
non-interacting Fermion models and interacting Fermion models. 

In an optimization process in the model space, 
the density is searched in a $v$-representable subset. 
Thus, in a realization of MR-DFT, 
we never meet the accumulation of unreachable crossing points, 
when we see a convergence of the density in a searching step. 
The existence of an $\varepsilon$-vicinity in the model space 
suggests that we have a physically converged model, 
which may continuously connect to the true electron system. 
Even when the original Kohn-Sham model is separated from 
the electron system by crossing points, 
the generalized Kohn-Sham model in MR-DFT can be settled in 
the $\varepsilon$-vicinity. 

\subsection*{Acknowledgement}
This work was supported by 
the Global COE Program (Core Research and Engineering of Advanced 
Material-Interdisciplinary Education Center for Materials Science), 
MEXT, Japan, 
Grand Challenges in next-generation integrated nanoscience, 
Grant-in-Aid for Scientific Research in 
Priority Areas (No. 17064006, No. 19051016) and 
a Grants-in-Aid for Scientific Research (No. 19310094).

\end{document}